\documentstyle[12pt]{article}

\def\beq{\begin{equation}} \def\eeq{\end{equation}}
\def\bea{\begin{eqnarray}} \def\eea{\end{eqnarray}}
\let\nn=\nonumber
\def\beann{\begin{eqnarray*}} \def\eeann{\end{eqnarray*}}

\let\a=\alpha \let\be=\beta \let\g=\gamma 
  \let\h=\eta 
\let\dh=\vartheta  \let\la=\lambda \let\m=\mu
\let\n=\nu  \let\p=\pi  \let\s=\sigma
 \let\ps=\psi
\let\ph=\varphi   \let\Ps=\Psi
  
 \let\G=\Gamma 

\let\qd=\quad \let\qqd=\qquad 

\def\tst#1{{\textstyle #1}}

\def\0{\over } \def\1{\vec }     \def\2{{1\over2}} \def\4{{1\over4}}
\def\5{\bar }  \def\6{\partial } \def\7#1{{#1}\llap{/}}

\def\<{\langle } \def\>{\rangle }

\let\auf=\uparrow \let\ab=\downarrow

  \def\CO{{\cal O}}
\def\i{{\rm i}} \def\tr{\mbox{tr}}

\renewcommand{\det}{\mbox{det}}

 \def\ch{\mbox{ch}}

\def\sign{\mbox{sign}}

\def\ssign{\mbox{\scriptsize sign}}

\def\sch{\mbox{\scriptsize ch}}

\begin{document}

\thispagestyle{empty}
\begin{center}
{\Large {\bf Correlations in the impenetrable electron gas\\}}
\vspace{7mm}
{\large F.~G\"{o}hmann$^\dagger$\footnote
{e-mail: goehmann@insti.physics.sunysb.edu},
A.~R.~Its$^\ddagger$\footnote
{e-mail: itsa@math.iupui.edu},
V.~E.~Korepin$^\dagger$\footnote
{e-mail: korepin@insti.physics.sunysb.edu}}\\
\vspace{5mm}
$^\dagger$Institute for Theoretical Physics,\\ State University of New
York at Stony Brook,\\ Stony Brook, NY 11794-3840, USA\\[1mm]
$^\ddagger$Department of Mathematical Sciences,\\ Indiana University --
Purdue University at Indianapolis (IUPUI),\\ 402 N. Blackford St.\\
Indianapolis, IN 46202-3216\\
\vspace{20mm}

{\large {\bf Abstract}}
\end{center}
\begin{list}{}{\addtolength{\rightmargin}{10mm}
               \addtolength{\topsep}{-5mm}}
\item
We consider non-relativistic electrons in one dimension with infinitely
strong repulsive delta function interaction. We calculate the long-time,
large-distance asymptotics of field-field correlators in the gas phase.
The gas phase at low temperatures is characterized by the ideal gas law.
We calculate the exponential decay, the power law corrections and the
constant factor of the asymptotics. Our results are valid at any
temperature. They simplify at low temperatures, where they are easily
recognized as products of free fermionic correlation functions with
corrections arising due to the interaction.
\\[2ex]
\end{list}

\clearpage

\subsection*{Introduction}
Non-relativistic electrons in one dimension, interacting strictly
locally through a delta potential constitute one of the
important exactly solvable many-body systems with applications in solid
state physics. Historically this was the first system to be solved
exactly by nested Bethe ansatz \cite{Yang67,Gaudin67}. Soon after an
exact formulation of its thermodynamics was obtained \cite{Takahashi71}.
In this article we address the problem of the exact calculation of
its correlation functions. Our work is based on a recently derived
determinant representation for the field-field correlators
\cite{IzPr97,IzPr98} valid in the case of infinite repulsion. The
determinant representation gives the field-field correlators as
functions of space and time variables $x$ and $t$ in a grand canonical
description, i.e.\ the correlators depend on temperature $T$, chemical
potential $\m$ and on the external magnetic field $B$. We study 
the the long-time, large-distance asymptotics of the field-field
correlators. We shall consider the limit $t \rightarrow \infty$
for fixed ratio $k_0 = x/2t$.

Let us write the Hamiltonian in terms of canonical anticommuting
quantum fields $\ps_\a (x,t)$ ($\a = \auf, \ab$),
\beq \label{ham}
     H = \int_{- \infty}^\infty dx \left\{
         \6_x \ps_\a^\dagger \6_x \ps_\a
	 + c : \! \left( \ps_\a^\dagger \ps_\a \right)^2 \! :
	 - \, \m \, \ps_\a^\dagger \ps_\a + B (\ps^\dagger \s^z \ps)
	 \right\} \qd.
\eeq
Here $\s^z$ is a Pauli matrix, and $c$ is the coupling constant. In
the following we shall consider infinite repulsion $c = + \infty$.
Then there is an explicit expression for the pressure as a function
of temperature, chemical potential and magnetic field
\cite{Takahashi71},
\beq \label{press}
     P = \frac{T}{2\p} \int_{- \infty}^\infty dk
         \ln \left( 1 + e^{(\m + B - k^2)/T} + e^{(\m - B - k^2)/T}
	     \right) \qd,
\eeq
which may serve as thermodynamic potential. Note that this expression
is formally the same as for a free Fermi gas with effective chemical
potential $\be T = \m + T \ln(2 \ch (B/T))$. Hence there are two
different zero temperature phases depending on whether $\lim_{T
\rightarrow 0+} \be T = \m + |B|$ is positive or negative. If
$\m + |B| > 0$, the density $D = \6 P/\6 \m$ has a positive limit as
the temperature goes to zero. If $\m + |B| < 0$, the density at zero
temperature vanishes. This is the case we are interested in. We shall
assume throughout the remainder of this article that $\be < 0$.
We call the phase determined by this condition the gas phase, since in
this phase pressure and density at low temperatures are related by the
ideal gas law,
\beq
     P = D T \qd.
\eeq
As the temperature increases the density becomes large, and
the ideal gas law does not hold anymore. There is a macroscopic number
of electrons in the system, which may carry a current. At strictly zero
temperature there are no particles in the system, and the system is an
insulator.

We shall consider the following two-point correlation functions of
local fields,
\bea \label{gp1}
     G_{\a \be}^+ (x,t) & = & \frac{\tr \left( e^{- H/T} \,
        \ps_\a (x,t) \ps_\be^\dagger (0,0) \right)}
	{\tr \left( e^{- H/T} \right)} \qd, \\ \label{gm1}
     G_{\a \be}^- (x,t) & = & \frac{\tr \left( e^{- H/T} \,
        \ps_\a^\dagger (x,t) \ps_\be (0,0) \right)}
	{\tr \left( e^{- H/T} \right)} \qd,
\eea
where $\a, \be = \auf, \ab$. Due to conservation of the total spin
$G_{\auf \ab}^\pm (x,t) = G_{\ab \auf}^\pm (x,t) = 0$. Furthermore,
due to the invariance of the Hamiltonian (\ref{ham}) under the
transformation $\auf \: \rightleftharpoons \, \ab$, $B
\rightleftharpoons - B$, we have the identity $G_{\ab \ab}^\pm (x,t)
= G_{\auf \auf}^\pm (x,t) |_{B \rightleftharpoons - B}$.
Hence, we can restrict our attention to the correlation functions
$G_{\auf \auf}^\pm (x,t)$.
\subsection*{Determinant representation}
Let us recall the determinant representation for the correlation
functions $G_{\auf \auf}^\pm (x,t)$, which was derived in
\cite{IzPr97,IzPr98}. We shall basically follow the account of
\cite{GIKP98}. Yet, it turns out to be very useful for further
calculations to rescale the variables and the correlation functions.
The rescaling
\bea \label{xtr}
     && x_r = - \sqrt{T} x /2 \qd, \qd
        t_r = T t /2 \qd, \\
     && g^\pm = G_{\auf \auf}^\pm /\sqrt{T} \qd, \\
     && h = B/T
\eea
removes the explicit temperature dependence from all expressions.
Furthermore, it will allow us to make close contact with results which
were obtained for the impenetrable Bose gas \cite{IIKS90,IIKV92,KBIBo}.
The index `$r$' in (\ref{xtr}) stands for `rescaled'. For the sake of
simplicity we shall suppress this index in the following three sections.
We shall come back to physical space and time variables only in the
last section, where we consider the low temperature limit.

The rescaled correlation functions $g^+$ and $g^-$ in the rescaled
variables can be expressed as \cite{GIKP98},
\bea \label{gp2}
     g^+ (x,t) & = &
        \frac{- e^{2 \i t (\be - h - \ln(2\sch (h)))}}{2 \p}
        \int_{- \p}^\p \! d\h \, \frac{F(\g,\h)}{1 - \cos(\h)} \,
	b_{++} \det\left( \hat I + \hat V \right) \qd, \\
     \label{gm2}
     g^- (x,t) & = &
	\frac{e^{- 2 \i t (\be - h - \ln(2\sch (h)))}}{4 \p \g}
        \int_{- \p}^\p \! d\h \, F(\g,\h) B_{--}
	\det\left( \hat I + \hat V \right) \qd.
\eea
Here $\g$ and $F(\g,\h)$ are elementary functions,
\bea
     \g & = & 1 + e^{2h} \qd, \\
     F(\g,\h) & = & 1 + \frac{e^{\i \h}}{\g - e^{\i \h}} +
                    \frac{e^{- \i \h}}{\g - e^{- \i \h}} \qd.
\eea
$\det(\hat I + \hat V)$ is the Fredholm determinant of the integral
operator $\hat I + \hat V$, where $\hat I$ is the identity operator,
and $\hat V$ is defined by its kernel $V(\la,\m)$. $\la$ and $\m$ are
complex variables, and the path of integration is the real axis. In
order to define $V(\la,\m)$ we have to introduce certain auxiliary
functions. Let us define
\bea
     \tau (\la) & = & \i(\la^2 t + \la x) \qd, \\
     \dh(\la) & = & \frac{1}{1 + e^{\la^2 - \be}} \qd, \\
         \label{eoflambda}
     E(\la) & = & \mbox{p.v.} \int_{- \infty}^\infty \! d\m \,
           \frac{e^{- 2 \tau(\m)}}{\p (\m - \la)} \qd, \\ \label{em}
     e_-(\la) & = &
        \sqrt{\frac{\dh(\la)}{\p}} e^{\tau(\la)} \qd, \\ \label{ep}
     e_+(\la) & = &
        \frac{1}{2} \sqrt{\frac{\dh(\la)}{\p}} e^{-\tau(\la)}
           \left\{ (1 - \cos(\h)) e^{2 \tau(\la)} E(\la) + \sin(\h)
	      \right\} \qd.
\eea
Note that $\dh(\la)$ is the Fermi weight. $V(\la,\m)$ can be expressed
in terms of $e_+$ and $e_-$,
\beq
     V(\la,\m) = \frac{e_+(\la) e_-(\m) - e_+(\m) e_-(\la)}{\la - \m}
        \qd.
\eeq

Denote the resolvent of $\hat V$ by $\hat R$,
\beq
     \left( \hat I + \hat V \right) \left( \hat I - \hat R \right) =
     \left( \hat I - \hat R \right) \left( \hat I + \hat V \right)
        = \hat I \qd.
\eeq
Then $\hat R$ is an integral operator with symmetric kernel,
\beq \label{kerr}
     R(\la, \m) = \frac{f_+(\la) f_-(\m) - f_+(\m) f_-(\la)}{\la - \m}
        \qd,
\eeq
which is of the same form as $V(\la,\m)$. The functions $f_\pm$ are
obtained as the solutions of the integral equations
\beq \label{fies}
     f_\pm(\la) + \int_{-\infty}^\infty \! d\m \, V(\la, \m) f_\pm(\m)
        = e_\pm (\la) \qd.
\eeq

We may now define the potentials
\beq \label{defpot}
     B_{ab} = \int_{- \infty}^\infty \! d\la \, e_a(\la) f_b(\la) \qd,
              \qd
     C_{ab} = \int_{- \infty}^\infty \! d\la \, \la e_a(\la) f_b(\la)
\eeq
for $a, b = \pm$. $B_{--}$ enters the definition of $g^- (x,t)$,
equation (\ref{gm2}). $b_{++}$ in (\ref{gp2}) is defined as
\beq
     b_{++} = B_{++} - G(x,t) \qd,
\eeq
where
\beq
     G(x,t) = \frac{1 - \cos(\h)}{2\p} \int_{- \infty}^\infty
	          \! d\la \, e^{- 2\tau(\la)} \\
            = \frac{(1 - \cos(\h)) e^{- \i \p /4}}{2 \sqrt{2 \p t}}
	          e^{\i x^2/2t} \qd.
\eeq
The remaining potentials $B_{ab}$ and $C_{ab}$ will be needed later.

It is instructive to compare the determinant representation (\ref{gp2})
for the correlation function $g^+ (x,t)$ with the corresponding
expression for impenetrable Bosons (cf e.g.\ page 345 of \cite{KBIBo}).
The main formal differences are the occurrence of the $\h$-integral in
(\ref{gp2}) and the occurrence of $\h$ in the definition of $e_+$. As
can be seen from the derivation of (\ref{gp2}) in \cite{IzPr98}, the
$\h$-integration is related to the spin degrees of freedom. For
$\h = \pm \p$ the expression $- \2 e^{2 \i \be t} b_{++}
\det(\hat I + \hat V)$ agrees with the field-field correlator for
impenetrable Bosons (remember, however, the different physical meaning
of~$\be$).

\subsection*{Differential equations}
As in case of impenetrable Bosons \cite{IIKS90,KBIBo} it is possible
to derive a set of integrable nonlinear partial differential equations
for the potentials $b_{++}$ and $B_{--}$ and to express the logarithmic
derivatives of the Fredholm determinant $\det(\hat I + \hat V)$ in
terms of the potentials $B_{ab}$ and $C_{ab}$.

The functions $f_\pm$ satisfy linear differential equations with respect
to the variables $x$, $t$, and $\be$,
\beq \label{lindiff}
     \hat L {f_+ \choose f_-} = \hat M {f_+ \choose f_-} =
        \hat N {f_+ \choose f_-} = 0 \qd.
\eeq
The Lax operators $\hat L$, $\hat M$ and $\hat N$ are given as
\bea
     \hat L & = & \6_x + \i \la \s^z - 2 \i Q \qd, \\
     \hat M & = & \6_t + \i \la^2 \s^z - 2 \i \la Q + \6_x U \qd, \\
     \hat N & = & 2 \la \6_\be + \6_\la + 2 \i t \la \s^z + \i x \s^z
                  - 4 \i t Q - 2 \6_\be U \qd,
\eea
where the matrices $Q$ and $U$ are defined according to
\beq
     Q = \left(\begin{array}{cc} 0 & b_{++} \\ B_{--} & 0
         \end{array} \right) \qd, \qd
     U = \left(\begin{array}{cc} - B_{+-} & b_{++} \\ - B_{--} & B_{+-}
         \end{array} \right) \qd.
\eeq

Mutual compatibility of the linear differential equations 
(\ref{lindiff}) leads to a set of nonlinear partial differential
equations for the potentials $b_{++}$ and $B_{--}$. Instead of using
$b_{++}$ and $B_{--}$, it is convenient to introduce the following
notations (cf.\ p.\ 352 of the book \cite{KBIBo}),
\bea
     g_+ = e^{2 \i t \be} b_{++} \qd &, & \qd
     g_- = e^{-2 \i t \be} B_{--} \qd, \\
     n = g_+ g_- = b_{++} B_{--} \qd &, & \qd
     p = g_- \6_x g_+ - g_+ \6_x g_- \qd.
\eea
The space and time evolution is driven by the separated nonlinear
Schr\"o\-dinger equation,
\bea \label{sepnls1}
     - \i \6_t g_+ & = & 2 \be g_+ + \tst{\2} \6_x^2 g_+ + 4 g_+^2 g_-
        \qd, \\ \label{sepnls2}
     \i \6_t g_- & = & 2 \be g_- + \tst{\2} \6_x^2 g_- + 4 g_-^2 g_+
        \qd.
\eea
$p$ and $n$ satisfy the continuity equation
\beq
     - 2 \i \6_t n = \6_x p \qd.
\eeq
The equations containing derivatives with respect to $\be$ are
\bea
     - \i \6_t \ph + 4 \6_\be p & = & 0 \qd, \\
     \6_x \ph + 8 \6_\be n + 2 & = & 0 \qd.
\eea
Here the function $\ph$ is defined as
\beq
     \ph (x,t,\be) = \frac{\6_\be \6_x g_+}{g_+} =
                     \frac{\6_\be \6_x g_-}{g_-} \qd.
\eeq

To describe the correlation functions (\ref{gp2}) and (\ref{gm2})
one has to relate the Fredholm determinant $\det(\hat I + \hat V)$
and the potentials $B_{ab}$ and $C_{ab}$. Let us use the abbreviation
$\s (x,t,\be) = \ln \det(\hat I + \hat V)$. The logarithmic derivatives
of the Fredholm determinant with respect to $x$, $t$ and $\be$ are
\bea \label{logdetx}
     \6_x \s & = & - 2\i B_{+-} \qd, \\ \label{logdett}
     \6_t \s & = & - 2\i (C_{+-} + C_{-+} + G(x,t) B_{--}) \qd, \\
     \6_\be \s & = & - 2\i t \6_\be (C_{+-} + C_{-+} + G(x,t) B_{--})
                     - 2\i x \6_\be B_{+-} - 2(\6_\be B_{+-})^2
		     \nn \\ \label{logdetb}
               &&    - 2\i t (B_{--} \6_\be b_{++}
		     - b_{++} \6_\be B_{--})
	             + 2(\6_\be b_{++})(\6_\be B_{--}) \qd.
\eea
For the calculation of the asymptotics of the Fredholm determinant
we further need the second derivatives of $\s$ with respect to space
and time,
\bea \label{logdetxx}
     \6_x^2 \s & = & 4 B_{--} b_{++} \qd, \\ \label{logdetxt}
     \6_x \6_t \s & = & 2\i (B_{--} \6_x b_{++} - b_{++} \6_x B_{--})
                        \qd, \\ \label{logdettt}
     \6_t^2 \s & = & 2\i (B_{--} \6_t b_{++} - b_{++} \6_t B_{--})
                     + 8 B_{--}^2 b_{++}^2 + 2 (\6_x B_{--})
		     (\6_x b_{++}) . \qd
\eea
Note that
\beq \label{limdhb}
     \lim_{\be \rightarrow - \infty} \s = 0 \qd.
\eeq
This follows from $\lim_{\be \rightarrow - \infty} \dh (\la) = 0$
and is important for fixing the integration constant in the calculation
of the asymptotics of the determinant.

The remarkable result of this section is that the differential
equations for $b_{++}$ and $B_{--}$ as well as the expressions
(\ref{logdetx})-(\ref{logdetb}) for the logarithmic derivative
of the Fredholm determinant are of the same form as in case
of impenetrable Bosons \cite{IIKS90,KBIBo}.
\subsection*{Asymptotics of the correlation functions}
The differential equations (\ref{sepnls1}), (\ref{sepnls2}) have
many solutions. $b_{++}$ and $B_{--}$ are the particular solution
of (\ref{sepnls1}) and (\ref{sepnls2}) which is fixed by the integral
equations (\ref{fies}). Alternatively $b_{++}$ and $B_{--}$ can be
obtained by solving a Riemann-Hilbert problem, which was formulated
in \cite{GIKP98}. The Riemann-Hilbert problem is the appropriate
tool to obtain the leading asymptotics of $b_{++}$, $B_{--}$, $\6_x \s$
and $\6_t \s$. In fact, using the method first suggested by Manakov
\cite{Manakov74} and further developed in \cite{Its81} and in
\cite{IIKV92}, it was shown in \cite{GIKP98} that the following
estimates hold,
\beq \label{bestimate}
     b_{++} = \CO \left(t^{- \2}\right) \qd, \qd
     B_{--} = \CO \left(t^{- \2}\right) \qd,
\eeq
and
\bea \label{dxsigma}
     \6_x \s & = & \frac{1}{\p} \int_{- \infty}^\infty \! d \la \;
                   \sign(\la - \la_0) \ln(\ph(\la,\be))
		   + \CO \left(t^{- \2}\right) \qd, \\
		   \label{dtsigma}
     \6_t \s & = & \frac{2}{\p} \int_{- \infty}^\infty \! d \la \;
                   \la \, \sign(\la - \la_0) \ln(\ph(\la,\be))
		   + \CO \left(t^{- \2}\right) \qd.
\eea
The function $\ph(\la,\be)$ in the above expressions is defined as
\beq \label{ph}
     \ph(\la,\be) = 1 + \dh(\la)\left(
                    e^{- \i \h \ssign(\la - \la_0)} - 1 \right) \qd.
\eeq
$\la_0$ is the stationary point of the phase $\tau(\la)$, $\tau'(\la_0)
= 0$, i.e.
\beq \label{sp}
     \la_0 = - \frac{x}{2t} \qd.
\eeq

Equations (\ref{bestimate})-(\ref{sp}) came out of the direct
asymptotic analysis of the Riemann-Hilbert problem. As we
shall explain in the following these equations, together with
(\ref{logdetx})-(\ref{logdettt}), fix the complete asymptotic expansion
of the correlation functions $g^+ (x,t)$ and $g^- (x,t)$ up to a single
unknown function.

It follows from (\ref{bestimate}) that $b_{++}$ and $B_{--}$ are a
decaying solution of the separated nonlinear Schr\"odinger equation
(\ref{sepnls1}), (\ref{sepnls2}). The form of the complete asymptotic
decomposition of the decaying solutions of the separated nonlinear
Schr\"odinger equation is known \cite{SeAb76,AbSeBo},
\bea \label{bppexp}
     b_{++} & = & t^{- \2} \left( u_0 + \sum_{n=1}^\infty
                  \sum_{k=0}^{2n} \frac{\ln^k 4t}{t^n} \, u_{nk} \right)
		  \exp \left\{ \frac{\i x^2}{2t} - \i \n \ln 4t \right\}
		  \qd, \\ \label{bmmexp}
     B_{--} & = & t^{- \2} \left( v_0 + \sum_{n=1}^\infty
                  \sum_{k=0}^{2n} \frac{\ln^k 4t}{t^n} \, v_{nk} \right)
		  \exp \left\{ - \frac{\i x^2}{2t} + \i \n \ln 4t
		  \right\} \qd.
\eea
Here the quantities $u_0$, $v_0$, $u_{nk}$, $v_{nk}$ and $\n$ are
functions of $\la_0 = -x/2t$ and of $\be$ and $\h$. Inserting the
above asymptotic expansions for $B_{--}$ and $b_{++}$ into the
differential equations (\ref{sepnls1}), (\ref{sepnls2}) we obtain
expressions for $u_{nk}$, $v_{nk}$ and $\n$ in terms of $u_0$ and $v_0$.
In other words, the two unknown functions $u_0$ and $v_0$ determine
the whole asymptotic expansion (\ref{bppexp}), (\ref{bmmexp}).
In particular,
\beq \label{co1}
     u_0 v_0 = - \frac{\n}{4} \qd.
\eeq
This relation can be used to calculate $\n$. (\ref{logdetxx}),
(\ref{bppexp}), (\ref{bmmexp}) and (\ref{co1}) imply that
\beq \label{findnu}
     \6_x^2 \s = - \frac{\n}{t} +
                 \CO \left( \frac{\ln^2 4t}{t^2} \right) \qd.
\eeq
Differentiating (\ref{dxsigma}) with respect to $x$ and comparing the
result with the right hand side of (\ref{findnu}) we find
\beq \label{nu}
     \n = - \frac{1}{2 \p} \ln \left( 1 - 2 (1 - \cos(\h)) \dh (\la_0)
              (1 - \dh (\la_0)) \right) \qd,
\eeq
such that the product $u_0 v_0$ is fixed, and we are indeed left with
only one unknown function, say, $u_0$.

The second logarithmic derivatives of the Fredholm determinant with
respect to space and time are given as functions of $b_{++}$ and
$B_{--}$ by the equations (\ref{logdetxx})-(\ref{logdettt}). Hence,
their complete asymptotics, too, is determined by the expansions
(\ref{bppexp}) and (\ref{bmmexp}). We may integrate (\ref{logdetxx})-%
(\ref{logdettt}) to obtain $\6_x \s$ and $\6_t \s$. The integration
constant, which is a function of $\be$, is fixed by the leading
asymptotics (\ref{dxsigma}), (\ref{dtsigma}). Then, integrating
(\ref{logdetx})-(\ref{logdetb}) yields $\s$ up to a numerical constant,
which follows from the asymptotic condition (\ref{limdhb}). The
calculation is the same as for the impenetrable Bose gas and can be
found on pages 455 - 457 of the book \cite{KBIBo}. It results in the
asymptotic formula
\bea \nn
     \s & = & \frac{1}{\p} \int_{-\infty}^\infty \! d \la \,
                 |x + 2 \la t| \ln(\ph(\la,\be))
		 + \frac{\n^2}{2} \ln 4t + \frac{\n^2}{2} \\ \nn
        &   & \qd + \, 2 \i \int_{- \infty}^\be \! d \be \,
	         (u_0 \6_\be v_0 - v_0 \6_\be u_0) \\ \nn
        &   & \qd + \, \frac{1}{2 \p^2} \int_{- \infty}^\be \! d \be \,
	         \left( \6_\be \int_{- \infty}^\infty \! d \la \,
		 \sign(\la - \la_0) \ln( \ph(\la,\be)) \right)^2 \\
		 \label{sigmaas}
        &   & \qd + \, \CO \left( \frac{\ln^4 4t}{t} \right) \qd.
\eea

So far the equations (\ref{bppexp}), (\ref{bmmexp}), (\ref{co1}),
(\ref{findnu}) and (\ref{sigmaas}) determine the asymptotics
of the correlation functions $g^+ (x,t)$ and $g^- (x,t)$ up to
an unknown function $u_0$. In order to calculate $u_0$ one has to
reexamine the Riemann-Hilbert problem. The asymptotic calculation
of \cite{GIKP98} has to be refined. For impenetrable Bosons a similar
refinement was achieved in \cite{IIKV92}. Fortunately, the result of
\cite{IIKV92} depends only on some general features of the functions
entering the Riemann-Hilbert problem in its canonical form. It turns
out that it also applies in our case. We can use equation (6.21) of
\cite{IIKV92} to obtain
\bea
     u_0 & = & \sqrt{\frac{\p}{2}} \;
                  \frac{e^{- \i \frac{3 \p}{4} + \i \Ps_1 -
		  \frac{\p \n}{2}}}{2 \dh (\la_0) \G ( - \i \n)} \\
         & = & \tst{\2} |\sin (\h/2)| \sqrt{\nu}
               \exp((\la_0^2 - \be)/2 + \i \Ps_0) \qd, \label{unull}
\eea
where
\bea \label{psinull}
     \Ps_0 & = & - \frac{3 \p}{4} + \mbox{arg} \, \G(\i \nu) + \Ps_1
                 \qd, \\ \label{psione}
     \Ps_1 & = & - \frac{1}{\p} \int_{- \infty}^\infty \!
	         d \la \; \sign(\la - \la_0) \ln |\la - \la_0|
	         \6_\la \ln(\ph(\la,\be)) \qd.
\eea
An alternative rigorous derivation of these formulae can be
performed via the Deift-Zhou nonlinear steepest descent method
\cite{DeZh94}. It is also worth comparing equations (\ref{unull})-%
(\ref{psione}) with the asymptotic formula for the solution of the
Cauchy problem of the NLS equation obtained by Zakharov and Manakov
\cite{ZaMa76} (see also \cite{SeAb76,AbSeBo} and \cite{DIZ93}).

Let us summarize our results. Inserting (\ref{bppexp}),
(\ref{bmmexp}), (\ref{sigmaas}) and (\ref{unull}) into (\ref{gp2})
and (\ref{gm2}) we obtain the following expressions for the leading
asymptotics of the correlation function,
\bea \label{gp}
     g^+ (x,t) & = &
	e^{\i x^2/2t + 2 \i t \be} e^{- 2 \i t (h + \ln(2\sch (h)))}
	\nn \\ && \qqd
        \int_{- \p}^\p \! d\h \, \frac{F(\g,\h)}{1 - \cos(\h)} \,
        C^+ (\la_0,\be,\h) \, (4t)^{\2 (\n - \i)^2} \, \nn \\ && \qqd
        \exp \left\{ \frac{1}{\p} \int_{-\infty}^\infty \! d \la \,
             |x + 2 \la t| \ln(\ph(\la,\be)) \right\} \qd, \\[2ex]
     \label{gm}
     g^- (x,t) & = &
	e^{- \i x^2/2t - 2 \i t \be} e^{2 \i t (h + \ln(2\sch (h)))}
	\nn \\ && \qqd
        \int_{- \p}^\p \! d\h \, \frac{F(\g,\h)}{2 \g} \,
        C^- (\la_0,\be,\h) \, (4t)^{\2 (\n + \i)^2} \, \nn \\ && \qqd
        \exp \left\{ \frac{1}{\p} \int_{-\infty}^\infty \! d \la \,
             |x + 2 \la t| \ln(\ph(\la,\be)) \right\} \qd,
\eea
where
\bea \label{cpp} \nn
     C^+ (\la_0,\be,\h) & = & - \, |\sin(\h/2)| \frac{\sqrt{\n}}{2 \p}
        \exp \left\{ \2 (\la_0^2 - \be) + \i \Ps_0 + \frac{\n^2}{2}
	\right. \\ && \qd
	- \int_{- \infty}^\be \! d \be \, (\i \nu /2 + \n \6_\be \Ps_0)
	\\ \nn && \qd
        \left. + \frac{1}{2 \p^2} \int_{- \infty}^\be \! d \be \,
	         \left( \6_\be \int_{- \infty}^\infty \! d \la \,
		 \sign(\la - \la_0) \ln( \ph(\la,\be)) \right)^2
		 \right\}, \qd \\[1ex] \label{cmm}
     C^- (\la_0,\be,\h) & = & C^+ (\la_0,\be,\h) \,
                              \exp(- (\la_0^2 - \be) - 2 \i \Ps_0)/
			      \sin^2 (\h/2) \qd.
\eea
These equations are valid for large $t$ and fixed finite ratio
$\la_0 = - x/2t$. Correlations in the pure space direction $t = 0$
will be discussed in a separate publication. We would like to emphasize
that (\ref{gp}) and (\ref{gm}) hold for arbitrary temperatures. Note
that there is no pole of the integrand at $\h = 0$, since $\sqrt{\n}
\sim |\h|$ for small $\h$ and thus $C^+ (\la_0,\be,\h) \sim \h^2$.
\subsection*{Asymptotics in the low temperature limit}
For the following steepest descent calculation we transform the
$\h$-integrals in (\ref{gp}), (\ref{gm}) into complex contour integrals
over the the unit circle, setting $z = e^{\i \h}$. Since we would like
to consider low temperatures, we have to restore the explicit
temperature dependence by scaling back to the physical space and time
variables $x$ and $t$ and to the physical correlation functions
$G_{\auf \auf}^\pm$. Recall that in the previous sections we have
suppressed an index `$r$' referring to `rescaled'. Let us restore this
index in order to define $k_0 = \la_0 \sqrt{T} = x/2t$, $\dh (k) =
\dh_r (k/\sqrt{T})$, $\ph (k,\be) = \ph_r (k/\sqrt{T},\be)$,
$C^\pm (k_0, \be, z) = C^\pm_r (\la_0, \be, \h)$, $F(\g, z) =
F_r (\g, \h)$. Then
\bea \label{gpc}
     G_{\auf \auf}^+ (x,t) & = & 2 \i \sqrt{T} \,
        e^{\i x^2/4t + \i t (\m - B)} \nn \\
	&& \oint dz \, \frac{F(\g,z)}{(z - 1)^2} C^+ (k_0,\be,z)
	   (2Tt)^{\2 (\n (z) - \i)^2} e^{t S(z)} \qd, \\ \label{gmc}
     G_{\auf \auf}^- (x,t) & = & - \i \sqrt{T} \,
        e^{- \i x^2/4t - \i t (\m - B)} \nn \\
	&& \oint dz \, \frac{F(\g,z)}{2 \g z} C^- (k_0,\be,z)
	   (2Tt)^{\2 (\n (z) + \i)^2} e^{t S(z)} \qd,
\eea
where
\beq
     S(z) = \frac{1}{\p} \int_{- \infty}^\infty \! dk \, |k - k_0|
            \ln(\ph(k,\be)) \qd.
\eeq

We would like to calculate the contour integrals (\ref{gpc}),
(\ref{gmc}) by the method of steepest descent. For this purpose we
have to consider the analytic properties of the integrands. Let us
assume that $k_0 \ge 0$, and let us cut the complex plane along the
real axis from $- \infty$ to $- e^{- \be}$ and from $- e^{\be -
k_0^2 /T}$ to $0$. The integrands in (\ref{gpc}) and (\ref{gmc})
can be analytically continued as functions of $z$ into the cut plane
with the only exception of the two simple poles of $F(\g,z)$ at
$z = \g^{\pm 1}$. We may therefore deform the contour of integration
as long as we never cross the cuts and take into account the pole
contributions, if we cross $z = \g$ or $z = \g^{-1}$.

The saddle point equation $\6 S/\6 z = 0$ can be represented in the
form
\beq \label{saddle}
     \int_0^\infty \frac{dk \, k}{1 + z^{-1} e^{- \be}
        e^{(k - k_0)^2 /T}} =
     \int_0^\infty \frac{dk \, k}{1 + z e^{- \be}
        e^{(k + k_0)^2 /T}} \qd.
\eeq
This equation was discussed in the appendix of \cite{GIKP98}. In
\cite{GIKP98} it was shown that (\ref{saddle}) has exactly one real
positive solution which is located in the interval $[0,1]$. It was
argued that this solution gives the leading saddle point contribution
to (\ref{gpc}) and (\ref{gmc}). At small temperatures (\ref{saddle})
can be solved explicitly. There are two solutions $z_\pm = \pm z_c$,
where
\beq \label{zc}
     z_c = \frac{T^{3/4}}{2 \p^{1/4} k_0^{3/2}} \, e^{- k_0^2/2T} \qd.
\eeq
In the derivation of (\ref{zc}) we assumed that $k_0 \ne 0$. The case
$k_0 = 0$ has to be treated separately (see below).

The phase $t S(z)$ has the low temperature approximation
\beq \label{lowphase}
     t S(z) = - 2 k_0 D t \left\{ \left( 1 - \frac{1}{z} \right)
	          z_c^2 + 1 - z \right\} \qd.
\eeq
Here $D = \6 P/\6 \m$ is the density of the electron gas. The low
temperature expansion (\ref{lowphase}) is valid in an annulus
$e^{\be - k_0^2/T} \ll |z| \ll e^{- \be}$, which lies in our cut plane.
The unit circle and the circle $|z| = z_c$ are inside this annulus.
We may thus first apply (\ref{lowphase}) and then deform the contour of
integration from the unit circle to the small circle $|z| = z_c$. Let
us parameterize the small circle as $z = z_c e^{\i \a}$, $\a \in
[- \p, \p]$. Then $S(z(\a)) = - 2k_0 D ((z_c - 1)^2 + 2z_c
(1 - \cos(\a)))$, which implies that the small circle is a steepest
descent contour and that on this contour $S(z_-) \le S(z) \le S(z_+)$.
The maximum of $S(z)$ on the steepest descent contour at $z = z_+$ is
unique and therefore provides the leading saddle point contribution to
(\ref{gpc}), (\ref{gmc}) as $t \rightarrow \infty$. The saddle point
approximation becomes good when $t S(z(\a)) = - 2k_0Dt((z_c - 1)^2 +
z_c \a^2 + \CO(\a^4))$ becomes sharply peaked around $\a = 0$. Hence,
the relevant parameter for the calculation of the asymptotics of
$G_{\auf \auf}^\pm$ is $2k_0Dt = xD$ rather than $t$. $xD$ has to be
large compared to $z_c^{-1}$. The parameter $xD$ has a simple
interpretation. It is the average number of particles in the interval
$[0,x]$. Let us consider two different limiting cases.
\begin{enumerate}
\item
$xD \rightarrow 0$, the number of electrons in the interval $[0,x]$
vanishes. In this regime the interaction of the electrons is
negligible. An electron propagates freely from 0 to $x$. $G_{\auf
\auf}^\pm$ cannot be calculated by the method of steepest descent. We
have to use the integral representation (\ref{gp}), (\ref{gm})
instead. Since $t S(z)$ and $\nu(z)$ tend to zero on the contour of
integration, the integrals in (\ref{gp}) and (\ref{gm}) are easily
calculated. We find
\bea \label{gpf}
     G_{\auf \auf}^+ (x,t) & = & \tst{\frac{e^{- \i \frac{\p}{4}}}
                                 {2 \sqrt{\p}}} \, t^{- \2}
				 e^{\i t (\m - B)} e^{\i x^2/4t} \qd,
				 \\ \label{gmf}
     G_{\auf \auf}^- (x,t) & = & \tst{\frac{e^{\i \frac{\p}{4}}}
                                 {2 \sqrt{\p}}} e^{(\m - B - k_0^2)/T}
				 \, t^{- \2} e^{- \i t (\m - B)}
				 e^{- \i x^2/4t} \qd.
\eea
As expected, equations (\ref{gpf}) and (\ref{gmf}) are the well known
result for free fermions.
\item
$xD \gg z_c^{-1}$, the average number of electrons in the interval
$[0,x]$ is large. This is the {\it true asymptotic region},
$x \rightarrow \infty$. In this region the interaction becomes
important. At the same time the method of steepest descent can be used
to calculate $G_{\auf \auf}^\pm$. This case will be studied below.
\end{enumerate}

In the process of deformation of the contour from the unit circle to
the small circle of radius $z_c$ we may cross the pole of the function
$F(\g,z)$ at $z = \g^{-1}$. Then we obtain a contribution of the pole
to the asymptotics of $G_{\auf \auf}^\pm$.  It turns out that the pole
contributes to $G_{\auf \auf}^\pm$, when the magnetic field is below a
critical positive value, $B_c = k_0^2/4$. Below this value the pole
contribution always dominates the contribution of the saddle point.
Hence, we have to distinguish two different asymptotic regions,
$B > B_c$ and $B < B_c$. On the other hand, if we consider the
asymptotics for fixed magnetic field, we have to treat the cases
$B > 0$ and $B \le 0$ separately. For $B > 0$ we have to distinguish
between a time like regime ($k_0^2 < 4B$) and a space like regime
($k_0^2 > 4B$). In these respective regimes we obtain the asymptotics:
\begin{description}
\item[Time like regime ($x < 4t \sqrt{B}$):]
\bea \label{gpas2}
     G_{\auf \auf}^+ (x,t) & = & \tst{\frac{e^{- \i \frac{\p}{4}}}
                                 {2 \sqrt{\p}}} \,
				 \tst{\frac{1}{\sqrt{4 \p z_c x D_\ab}}}
                                 \, t^{- \2 - \i \nu(z_c)}
                                 e^{\i t (\m - B)} e^{\i x^2/4t}
				 e^{- x D_\ab} \, , \\ \label{gmas2}
     G_{\auf \auf}^- (x,t) & = & \tst{\frac{e^{\i \frac{\p}{4}}}
                                 {2 \sqrt{\p}}} \,
				 \tst{\frac{e^{(\m - B - k_0^2)/T}}
				 {\sqrt{4 \p z_c x D_\ab}}}
                                 \, t^{- \2 + \i \nu(z_c)}
                                 e^{- \i t (\m - B)} e^{- \i x^2/4t}
			         e^{- x D_\ab} \, ,
\eea
\end{description}
where
\bea \label{nuzp}
     \nu(z_c) & = & - \; \frac{2D_\ab k_0^{3/2} e^{- k_0^2/2T}}
                          {\p^{1/4} T^{5/4}} \qd, \\
     D_\ab & = & \tst{\2 \sqrt{\frac{T}{\p}} e^{(\m + B)/T}} \qd.
\eea
$D_\ab = \6 P /\6 (\m + B)$ is the low temperature expression for the
density of down-spin electrons.
\begin{description}
\item[Space like regime ($x > 4t \sqrt{B}$):]
\bea \label{gpas}
     G_{\auf \auf}^+ (x,t) & = & \tst{\frac{e^{- \i \frac{\p}{4}}}
                                 {2 \sqrt{\p}}}
				 \, t^{- \2 - \i \nu(\g^{-1})}
                                 e^{\i t (\m - B)} e^{\i x^2/4t}
				 e^{- x D_\ab} \, , \\ \label{gmas}
     G_{\auf \auf}^- (x,t) & = & \tst{\frac{e^{\i \frac{\p}{4}}}
                                 {2 \sqrt{\p}}} \,
				 e^{(\m - B - k_0^2)/T}
                                 t^{- \2 + \i \nu(\g^{-1})}
                                 e^{- \i t (\m - B)} e^{- \i x^2/4t}
			         e^{- x D_\ab} \, ,
\eea
\end{description}
where
\beq \label{nugamma}
     \nu(\g^{-1}) = - \; \frac{e^{(3B + \m - k_0^2)/T}}{2\p} \qd.
\eeq
For $B \le 0$ there is no distinction between time and space like
regimes. The asymptotics is always given by the equations (\ref{gpas})
and (\ref{gmas}).

It is instructive to compare the asymptotic expressions (\ref{gpas2})-%
(\ref{nugamma}) for the correlation functions with the corresponding
asymptotics for free fermions, (\ref{gpf}), (\ref{gmf}). We can
interpret the factors $ t^{- \2 \pm \i \nu(z_+)} e^{- x D_\ab}/
\sqrt{4 \p z_c x D_\ab}$ in (\ref{gpas2}) and (\ref{gmas2}) and
$t^{\pm \i \nu(\g^{-1})} e^{- x D_\ab}$ in (\ref{gpas}) and (\ref{gmas})
as low temperature corrections to the free fermionic correlation
functions. The occurrence of the density of down-spin electrons in the
exponential factors in equations (\ref{gpas2})-(\ref{nugamma}) has a
natural interpretation. Correlations decay due to interaction. Because
of the Pauli principle and the locality of the interaction in our
specific model, equation (\ref{ham}), up-spin electrons only interact
with down-spin electrons. Therefore the correlation length is expected
to be a decreasing function of the density of down-spin electrons,
which diverges as the density of down spin electrons goes to zero.
We see from (\ref{gpas2})-(\ref{nugamma}) that the low temperature
expression for the correlation length is just $1/D_\ab$ and thus meets
these expectations.

In the limit $B \rightarrow - \infty$, $\m \rightarrow - \infty$,
$\m - B$ fixed there are no $\ab$-spin electrons left in the system,
$D_\ab \rightarrow 0$. This is the free fermion limit. In the free
fermion limit $B < B_c$, and the asymptotics of $G_{\auf \auf}^+ (x,t)$
and $G_{\auf \auf}^- (x,t)$ are given by equations (\ref{gpas}),
(\ref{gmas}), which turn into the expressions (\ref{gpf}), (\ref{gmf})
for free fermions.

The pure time direction $k_0 = 0$ requires a separate calculation.
For $k_0 = 0$ the saddle point equation (\ref{saddle}) has the solutions
$z = \pm 1$ for all temperatures. The unit circle is a steepest descent
contour with unique maximum of $S(z)$ at $z = 1$, which gives the
leading asymptotic contribution to the integrals in (\ref{gpc}) and
(\ref{gmc}). We find algebraically decaying correlations,
\beq
     G_{\auf \auf}^+ (0,t) = C_0^+ t^{-1} e^{\i t (\m - B)} \qd, \qd
     G_{\auf \auf}^- (0,t) = C_0^- t^{-1} e^{- \i t (\m - B)} \qd,
\eeq
where
\bea
     C_0^+ & = & \frac{e^{- \i \frac{\p}{4}}}
                 {2 \sqrt{2 \p T}}
                 (1 + 2 e^{- 2B/T}) \nn \\ &&
		 \left[ (e^{(\m + B)/T} + e^{(\m - B)/T})
		 (1 + e^{(\m + B)/T} + e^{(\m - B)/T}) \right]^{- \2}
		 \qd, \\
     C_0^- & = & \frac{e^{\i \frac{\p}{4}}}
                 {2 \sqrt{2 \p T}} \,
                 \frac{1 + 2 e^{- 2B/T}}{1 + e^{2B/T}}
		 \left[ \frac{e^{(\m + B)/T} + e^{(\m - B)/T}}{
		 1 + e^{(\m + B)/T} + e^{(\m - B)/T}} \right]^\2
		 \qd.
\eea
These formulae are valid at any temperature.
\subsection*{Acknowledgement}
This work was partially supported by the National Science Foundation
under grants number PHY-9605226 (V.K.) and number DMS-9801608 (A.I.),
and by the Deutsche Forschungsgemeinschaft under grant number
Go 825/2-1 (F.G.).


\begin{thebibliography}{10}

\bibitem{Yang67}
C.~N. Yang, Phys. Rev. Lett. {\bf 19}, 1312 (1967).

\bibitem{Gaudin67}
M.~Gaudin, Phys. Lett. A {\bf 24}, 55 (1967).

\bibitem{Takahashi71}
M.~Takahashi, Prog. Theor. Phys. {\bf 46}, 1388 (1971).

\bibitem{IzPr97}
A.~G. Izergin and A.~G. Pronko, Phys. Lett. A {\bf 236}, 445 (1997).

\bibitem{IzPr98}
A.~G. Izergin and A.~G. Pronko,
`Temperature correlators in the two-component one-dimensional gas',
PDMI preprint 19/1997, solv-int/9801004 (1998).

\bibitem{GIKP98}
F.~G\"ohmann, A.~G. Izergin, V.~E. Korepin, and A.~G. Pronko,
`Time and temperature dependent correlation functions of the
1D impenetrable electron gas', ITP-SB-98-34, cond-mat/9805192 (1998).

\bibitem{IIKS90}
A.~R. Its, A.~G. Izergin, V.~E. Korepin, and N.~Slavnov,
Int. J. Mod. Phys. B {\bf 4}, 1003 (1990).

\bibitem{IIKV92}
A.~R. Its, A.~G. Izergin, V.~E. Korepin, and G.~G. Varzugin,
Physica D {\bf 54}, 351 (1992).

\bibitem{KBIBo}
V.~E. Korepin, N.~M. Bogoliubov, and A.~G. Izergin, {\em Quantum
Inverse Scattering Method and Correlation Functions},
Cambridge University Press, (1993).

\bibitem{Manakov74}
S.~V. Manakov, Sov. Phys. JETP {\bf 38}, 693 (1974).

\bibitem{Its81}
A.~R. Its, Sov. Math. Dokl. {\bf 24}, 452 (1981).

\bibitem{SeAb76}
H.~Segur and M.~J. Ablowitz, J. Math. Phys. {\bf 17}, 710 (1976).

\bibitem{AbSeBo}
M.~J. Ablowitz and H.~Segur, {\em Solitons and the Inverse Scattering
  Transform}, SIAM, Philadelphia,  (1981).

\bibitem{DeZh94}
P.~A. Deift and X.~Zhou, Comm. Math. Phys. {\bf 165}, 175 (1994).

\bibitem{ZaMa76}
V.~E. Zakharov and S.~V. Manakov, Sov. Phys. JETP {\bf 44}, 106 (1976).

\bibitem{DIZ93}
P.~A. Deift, A.~R. Its, and X.~Zhou, {\em Important Developments in
Soliton Theory}, page 181, Springer-Verlag,  (1993).

\end{thebibliography}

\end{document}